# New perspectives on asymmetric bending behavior: A lesson learned from leaves

Anran Wei [1], Zhenbin Guo [2], Fenglin Guo [1] *

[1] School of Naval Architecture, Ocean and Civil Engineering (State Key Laboratory of Ocean Engineering), Shanghai Jiao Tong University, Shanghai 200240, China

[2] Institute of Semiconductor Manufacturing Research, Shenzhen University, Shenzhen 518060, China

* Corresponding author, E-mail: *flguo@sjtu.edu.cn*

**Abstract**

Designing materials or structures that can achieve asymmetric shape-shifting in response to symmetrically switching stimuli is a promising approach to enhance the locomotion performance of soft actuators/robots. Inspired by the geometry of slender leaves of many plants, we find that the thin-walled beam with a U-shaped cross section exhibits asymmetric deformation behaviors under bending with opposite orientations. Although this novel mechanical property has been long noticed and utilized in some applications, its mechanism is still unclear so far. In this study, we attribute this asymmetric bending behavior of thin-walled U-shaped beams to the buckling of sidewalls caused by the bending-induced compressive effect. Based on the Euler-Bernoulli beam theory and Kirchhoff-Love thin plate theory, a simple but efficient model is established to derive the critical moment for the sidewall buckling in a semi-analytical form. Finite element analysis simulations and experiments are employed to validate the theoretical foundations of our findings. The results of our work not only shed light on the mechanics underlying the asymmetric bending behavior of thin-walled U-shaped beams, but also open up new avenues for the structure design of high-performance soft actuators/robots and other novel devices.

Recent huge advances in responsive materials have stimulated the development of soft robotics [1-5]. For most reported soft actuators/robots made from responsive materials, the external stimulus will trigger the bending deformation of structures [6-10] to generate controllable mechanical actuation for many applications, such as mass lifting [11-13] and transporting [14-16]. If the stimulus is switched symmetrically, for example, flipping the direction of a magnetic field, the bending deformation will be reversed and show mirror symmetry. By repeatedly switching the stimulus, the shape of structures will be transformed periodically, which is one of the major propulsion modes of soft robots [17-19]. However, the propulsion driven by the symmetric actuation usually results in low locomotion efficiency. That is why efficient locomotion is usually achieved by propulsions with asymmetric processes in the animal kingdom. For example, the wings of birds [20-22] or legs of frogs [23-25] show asymmetric shape-shifting and motion attitude in a propulsion cycle of flying or swimming, which determines the amount of net propulsion for moving forward. This phenomenon inspires us that designing structures with asymmetric bending behavior under actuations of symmetric stimuli is promising to improve the locomotion performance of soft robots. A few explorations have been made to validate this idea [26,27]. In a representative work by Wu et al. [26], an asymmetric joint structure is designed to break the bending symmetry, which brings efficient dynamic performance to a swimming soft robot.

Besides the wings or legs with delicate structures consisting of bond joints, ligaments and muscles, there is another much simpler structure in nature exhibiting the asymmetry shape-shifting behavior as well. As shown in Fig. 1a, many plants have long yet almost straight leaves with edges upward curled. The shape of these slender leaves can be abstracted into a thin-walled beam with a U-shaped cross section. In most situations, these leaves can hold their long-span geometries when subjected to the bending by the self-weight and other downward loading weights like raindrops or biofoulings. If these leaves are turned over to make the opening of cross section downward, their bending resistance to download loads will be significantly weakened. Take the leaves of *Chlorophytum Comosum* as a typical example. A straight part from a leaf of *Chlorophytum Comosum* is dissected and placed horizontally with one end fixed and the other end free to set a cantilever system. For the same weight applied on the free end, the leaf with cross section facing downward is bent more severely than itself facing upward, as shown in Fig. 1b. The observation in leaves inspires us that the thin-walled U-shaped beam could show an asymmetric bending behavior. This structure will bring a new opportunity in the design of soft actuators/robots with enhanced locomotion performance.

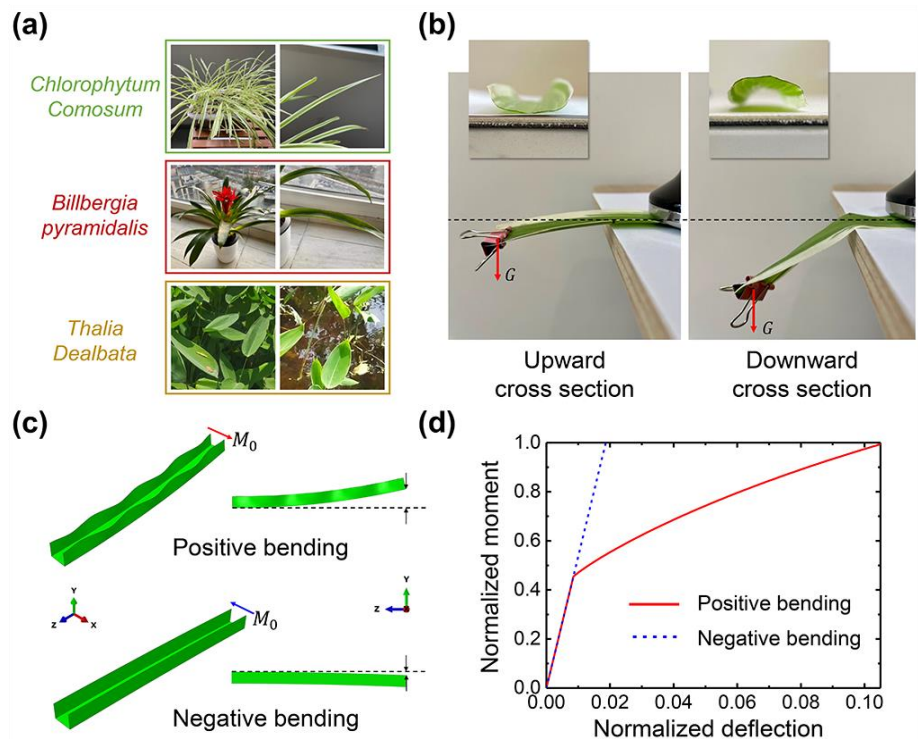

**Fig. 1 Asymmetric bending behavior of thin-walled U-shaped beams. (a)** Images of many slender leaves with a U-shaped cross section. **(b)** Bending deformation of a slender leaf of *Chlorophytum Comosum* with its U-shaped cross section facing upward and downward. For the two situations, one end

of the slender leaf is fixed and the other end is applied with the same loads. **(c)** Deformation analysis of thin-walled U-shaped beams under positive and negative bending by finite element analysis (FEA) simulations. The directions of bending moment following the right-hand rule are indicated. **(d)** Normalized moment-deflection curves for positive and negative bending obtained from FEA simulations.

This characteristic of thin-walled U-shaped beams has long been noticed and also utilized in other engineering scenarios. For example, the steel tape measure is manufactured to have a similar U-shaped cross section so that it can be easily folded from only one side. However, to the best of our knowledge, the mechanics underlying the asymmetric bending behavior remains unclear so far. Unveiling the related deformation mechanism can contribute greatly to the diverse applications of thin-walled U-shape beams, which is especially essential to the structure design of state-of-the-art soft actuators/robots applied under actuations of symmetric stimuli. Therefore, our study aims to provide theoretical foundations for an in-depth understanding of this novel mechanical behavior.

FEA simulations are carried out as virtual experiments for the analysis of the deformation details of thin-walled U-shaped beams under bending with opposite orientations. Beam models are constructed with one end clamped and the other end suffering a given moment. For this loading condition, the total moment in any cross section of a beam is a constant, which can be regarded as a pure bending. Reversing the direction of the moment will switch the bending orientations. Here, we define the overall bending that is towards the opening direction of the U-shaped cross section as *positive bending*, otherwise as *negative bending*, as illustrated in Fig. 1c. For the beam under positive bending, its sidewalls buckle when a small critical bending moment is reached, which generates a nearly periodic rippling except in the limited regions near the two ends. However, there is no such buckling of sidewalls appearing during the negative bending, even under a larger bending moment. Meanwhile, the moment-deflection curves for positive and negative bending are also obtained from FEA simulations, as plotted in Fig. 1d. The maximum moment is normalized to 1 and the deflection is normalized by the beam length. Initially, the two curves are coincided, showing the same increasing profiles. Then the increase for positive bending significantly slows down accompanied with the buckling of sidewalls, indicating the degeneration of bending resistance. On the contrary, the curve for negative bending keeps rising and the bending stiffness is steadily maintained. These FEA results confirm the asymmetric bending behaviors of thin-walled U-shaped beams. Combining the deformation analysis, we can conclude that such a novel mechanical property originates from the sidewall buckling only occurring under positive bending. Predicting the critical loads for the buckling is essential for triggering this novel mechanical property in practical applications. For the thorough understanding of the underlying mechanism, a mechanical model is thus required to interpret and predict the buckling phenomenon under positive bending.

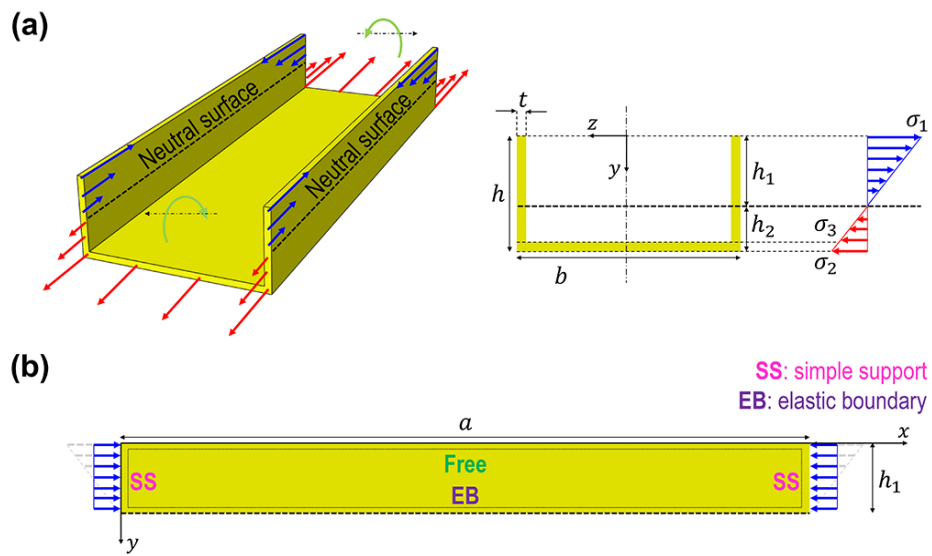

**Fig. 2 Schematics of the mechanical modeling on the thin-walled U-shaped beam under positive bending. (a)** Stress distribution in the cross section of beam. The structural geometries of beam (including the beam length $a$, height $h$, width $b$ and wall thickness $t$) and the positions of neutral

surface ($h_1$ and $h_2$) are indicated. **(b)** Thin plate model for modeling the buckling of sidewalls above the neutral surface under positive bending.

Consider a slender thin-walled U-shaped beam under pure bending with a length of $a$, a width of $b$, a height of $h$, and a uniform wall thickness of $t$, as illustrated in Fig. 2a. In this model, all deformation is within the elastic limit of material. When the positive bending is applied, compressive and tensile effects will be respectively produced above and below the neutral surface. If the neutral surface is located within the sidewalls, the position of neutral surface can be derived by the Euler-Bernoulli beam theory as

$$h_1 = \frac{2(h-t)^2 + b(2h-t)}{2b + 4(h-t)} \tag{1}$$

Due to the pre-condition of $h_1 \leq h - t$, a requirement on the geometries of cross section for the solution of $h_1$ can be derived as

$$2\left(\frac{h}{b} - \frac{t}{b}\right)\left(\frac{h}{t} - 1\right) \geq 1 \tag{2}$$

Since $t$ is much smaller than $h$ and $b$ for thin-walled beams, this requirement is satisfied in most cases as long as the height and width of U-shaped beams are comparable. For cases with $(h-t)/b \geq 0.5$, the relationship in Eq. (2) must hold for any cases. After knowing the position of neutral surface, the moment of inertia of the U-shaped cross section can be calculated as

$$I_z = \frac{bh^3}{12} + bh\left(h_1 - \frac{h}{2}\right)^2 - \frac{(b-2t)(h-t)^3}{12} - (b-2t)(h-t)\left(h_1 - \frac{h}{2} + \frac{t}{2}\right)^2 \tag{3}$$

The region above the neutral surface can be modeled by a thin plate under compression with a size of $a \times h_1$ and a thickness of $t$, as illustrated in Fig. 2b. It is known that thin plates are prone to buckle under compression. As the positive bending is aggravated, the increased compression will finally trigger the buckling of the thin plate, which interprets the unique rippling of sidewalls observed in the above FEA simulations. To simplify the theoretical derivation, the compressive stress with triangular distribution at the left and right boundaries of the plate is approximated as a uniformly distributed planar stress $P_x$ that is equivalent in the sense of resultant force. The Kirchhoff-Love thin plate theory tells that the buckling of a thin plate under uniform planar compression is governed by the equation of

$$D\nabla^4 w + P_x \frac{\partial^2 w}{\partial x^2} = 0 \qquad (4)$$

where $D$ is the bending stiffness of the thin plate and $w$ represents its out-of-plane deflection along the $z$-direction. The coordinate adopted in our model can be referred in Fig. 2. $D$ is a function of elastic modulus $E$, Poisson's ratio $v$ and plate thickness $t$, which follows the form of $D = Et^3/12(1 - v^2)$. For simplification, the left ($x = 0$) and right sides ($x = a$) of the plate are treated as two simply supported boundaries. To satisfy the left and right boundary conditions, $w = \sum_{m=1}^{\infty} Y_m(y)\sin(m\pi x/a)$ is taken as the general solution of Eq. (4), where $Y_m(y)$ is an undetermined function. Substituting it into Eq. (4) leads to an ordinary differential equation (ODE) of $Y_m(y)$ whose eigenvalues are $\pm\alpha$, $\pm\beta$ with $\alpha = \sqrt{m\pi(m\pi/a + \sqrt{P_x/D})/a}$ and $\beta = \sqrt{m\pi(m\pi/a - \sqrt{P_x/D})/a}$. Boundary conditions of the upper ($y = 0$) and lower sides ($y = h_1$) are required for solving the ODE. Free boundary condition with zero moment and force can be easily written for the upper side as

$$\begin{cases} \left(\frac{\partial^2 w}{\partial y^2} + v\frac{\partial^2 w}{\partial x^2}\right)\bigg|_{y=0} = 0 \\ \left[\frac{\partial^3 w}{\partial y^3} + (2 - v)\frac{\partial^3 w}{\partial x^2 \partial y}\right]\bigg|_{y=0} = 0 \end{cases} \qquad (5)$$

However, the constraints on the lower side are more complicated, since it is connected to the part below the neutral surface of which structural rigidity cannot be ignored. To handle this problem, we idealize the lower boundary of the plate suffering elastic constraints imposed by the structure below the neutral surface, just like being tied by an effective spring. That is, the shear force and moment in the lower side are modeled to be proportional to its out-of-plane displacement and rotation angle, respectively, as following forms of

$$\begin{cases} -D\left[\dfrac{\partial^3 w}{\partial y^3} + (2-v)\dfrac{\partial^3 w}{\partial x^2 \partial y}\right]\bigg|_{y=h_1} = -k_1 w|_{y=h_1} \\ -D\left(\dfrac{\partial^2 w}{\partial y^2} + v\dfrac{\partial^2 w}{\partial x^2}\right)\bigg|_{y=h_1} = k_2 \dfrac{\partial w}{\partial y}\bigg|_{y=h_1} \end{cases} \qquad (6)$$

Here $k_1$ and $k_2$ are two effective elastic coefficients with dimensions of N·m$^{-2}$ and N, respectively. Using the unit load method from the principle of virtual work, $k_1$ and $k_2$ can be approximately expressed as (See Supplemental Material [28] for detailed derivations)

$$\begin{cases} k_1 = \dfrac{Et^3}{4h_2^3 + 6h_2^2 b} \\ k_2 = \dfrac{Et^3}{12h_2 + 6b} \end{cases} \qquad (7)$$

Then the general solution of the ODE is substituted into these boundary conditions. The planar stress is then rewritten as $P_x = n\pi^2 D/h_1^2$ where $n$ is a nondimensional coefficient, and more nondimensional parameters are introduced including the aspect ratio $\zeta = a/h_1$, the normalized eigenvalues $\bar{\alpha} = a\alpha = \sqrt{m^2\pi^2 + m\pi^2\zeta\sqrt{n}}$ and $\bar{\beta} = a\beta = \sqrt{m^2\pi^2 - m\pi^2\zeta\sqrt{n}}$, as well as the normalized elastic coefficients $\bar{k}_1 = a^3 k_1/D$ and $\bar{k}_2 = ak_2/D$. To ensure the non-zero solutions, the determinant of the coefficient matrix should be equal to zero, which is written in a dimensionless form as

$$\begin{vmatrix} \delta_2\eta_3 - \delta_1\bar{k}_1 - \delta_4\dfrac{\eta_1\eta_4}{\eta_2} + \delta_3\dfrac{\eta_1}{\eta_2}\bar{k}_1 & \delta_1\eta_3 - \delta_2\bar{k}_1 - \delta_3\eta_3 + \delta_4\dfrac{\eta_3}{\eta_4}\bar{k}_1 \\ \delta_1\eta_1 - \delta_2\bar{k}_2\bar{\alpha} - \delta_3\eta_1 + \delta_4\dfrac{\eta_1}{\eta_2}\bar{k}_2\bar{\beta} & \delta_2\eta_1 - \delta_1\bar{k}_2\bar{\alpha} - \delta_4\dfrac{\eta_2\eta_3}{\eta_4} + \delta_3\dfrac{\eta_3}{\eta_4}\bar{k}_2\bar{\beta} \end{vmatrix} = 0 \qquad (8)$$

where $\delta_i$ and $\eta_i$ ($i = 1, 2, 3, 4$) are the combinations of these nondimensional parameters as

$$\begin{cases} \delta_1 = \cosh\dfrac{\bar{\alpha}}{\zeta}, \delta_2 = \sinh\dfrac{\bar{\alpha}}{\zeta} \\ \delta_3 = \cosh\dfrac{\bar{\beta}}{\zeta}, \delta_4 = \sinh\dfrac{\bar{\beta}}{\zeta} \end{cases} \text{ and } \begin{cases} \eta_1 = \bar{\alpha}^2 - m^2\pi^2 v \\ \eta_2 = \bar{\beta}^2 - m^2\pi^2 v \\ \eta_3 = \bar{\alpha}^3 - \bar{\alpha}m^2\pi^2(2-v) \\ \eta_4 = \bar{\beta}^3 - \bar{\beta}m^2\pi^2(2-v) \end{cases} \qquad (9)$$

Combining Eq. (8) and (9), we can numerically calculate the values of $n$ for arbitrary integers $m$. The minimum one $n_{\min}$ is taken for the final solution of critical load. Using the equivalence of the resultant force mentioned above and basic rules in Euler-Bernoulli beam theory, the critical moment $M_{\mathrm{cr}}$ for the sidewall buckling under positive bending is provided as

$$M_{\mathrm{cr}} = \frac{2n_{\min}\pi^2 DI_z}{h_1{}^3 t} \qquad (10)$$

The model of the bending-induced buckling has been established now, allowing us to investigate the effects of structural geometries of beam on the critical loads. To satisfy the condition of Eq. (2) for the position of neutral surface and the requirement of slender beam for the Euler-Bernoulli beam theory, cases with $(h-t)/b \geq 0.5$ and $h/a \leq 0.1$ are considered in the following discussion. Unless otherwise specified in main texts and figures, the variables $a$, $t$, $E$ and $v$ in these cases are taken as 600 mm, 1 mm, 70 GPa and 0.3, respectively.

We firstly study the dependence of $M_{\mathrm{cr}}$ on the wall thickness $t$. Numerical results by our model are plotted in Fig. 3a and 3b for U-shaped beams with different section sizes. It is observed that $M_{\mathrm{cr}}$ increases monotonically with $t$. Power function can be used to fit the relationship between $M_{\mathrm{cr}}$ and $t$. The exponents of best fittings are 2.95 and 2.97 for the two groups, respectively. It can be said that $M_{\mathrm{cr}}$ is almost a cubic function of $t$. The buckling analysis in FEA simulations gives very good validations on the theoretical predictions of $M_{\mathrm{cr}}$, indicating the effectiveness of our model in describing the sidewall buckling. Meanwhile, it is noticed from the comparison between Fig. 3a and 3b that U-shaped beams with the same wall thickness exhibit almost the same critical moments in spite of their different section sizes. The interesting finding stimulates more detailed examinations on the effects of $h$ and $b$ on $M_{\mathrm{cr}}$.

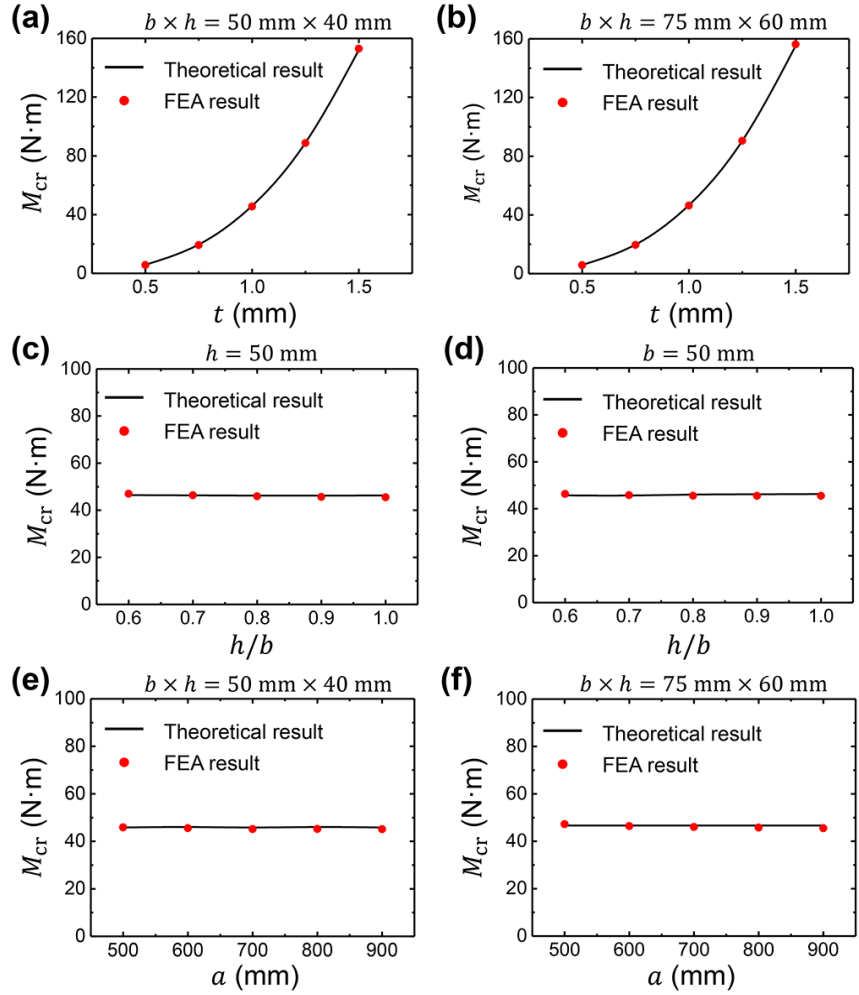

**Fig. 3 Effects of structural geometries of beam on the critical moment for the sidewall buckling under positive bending.** The dependences of critical moment $M_{cr}$ on the wall thickness $t$ **(a, b)**, beam height-width ratio $h/b$ **(c, d)**, beam length $a$ **(e, f)** are plotted. Theoretical and finite element analysis (FEA) results are shown by the black lines and red dots, respectively.

Fig. 3c and 3d shows the results of $M_{cr}$ for different ratios $h/b$ under fixed $h$ and $b$, respectively. We can also see that changing section sizes, whether the height $h$ or the width $b$ or the width-height ratio $h/b$, has a trivial influence on the critical moment. Our model and FEA simulations provide quantitatively closed results of $M_{cr}$, confirming this insensitivity on the section size. Moreover, the U-shaped beams with different lengths $a$ are also analyzed. Similar to the results for $h$ and $b$, $M_{cr}$ is almost independent of the $a$ as shown in Fig. 3e and 3f, which is also validated by FEA simulations. That gives us a guideline that adjusting the width, height and length may not be efficient to manipulate the triggering condition of the asymmetric bending behavior of thin-walled U-shaped beams.

The above theoretical modeling unveils the physics of the asymmetric bending behaviors but derives $M_{cr}$ in a complicated implicit form with many variables involved and couped. A simple and reliable explicit expression of $M_{cr}$ could be more favorable for engineering guidelines. As shown in the foregoing discussion, we have already proved that some of the variables such as $h$, $b$, $a$ and $v$ play inessential roles in $M_{cr}$, and $M_{cr}$ is nearly proportional to $t^3$ and $E$. For convenience, we can further simplify the semi-analytical solution of $M_{cr}$ by fitting an empirical formula where only the dominant variables are retained. The empirical formula of $M_{cr}$ follows the linear relationship of $M_{cr} \propto Et^3$ of which the proportional coefficient can be obtained from the fitting of several FEA results listed in Table S1 (See Supplemental Material [28] for detailed data). Notice that cases with the same values of $Et^3$ shows almost the same $M_{cr}$ in FEA simulations, supporting the functional form used in the empirical formula. The best fitting of $M_{cr}$ is written as

$$M_{cr} = 0.65Et^3 \qquad (11)$$

with very good linearity of $R^2 = 0.99$. To examine the accuracy of this simple empirical formula, three samples of thin-walled U-shaped beams are manufactured using 6063 aluminum alloy, which have the sizes of $h \times b \times t =$ 10 mm × 10 mm × 0.5 mm, 12 mm × 12 mm × 0.5 mm and 11 mm × 16 mm × 0.7 mm, respectively. To achieve the pure bending condition, the four-point bending test is conducted as the experimental settings shown in Fig. 4a. The distance between two supporting pins and loading pins are 150 mm and 75 mm, respectively. A larger portion of the sample between two loading pins is subjected to a constant bending moment, which can be converted from the force applied on the loading pins. The critical moments for the sidewall buckling are recorded and plotted in Fig. 4b, which agree well with the estimations using Eq. (11). The experimental results also indicate that the U-shaped beams with different section sizes but the same wall thickness show similar values of $M_{cr}$, validating the simplification of our model in Eq. (11).

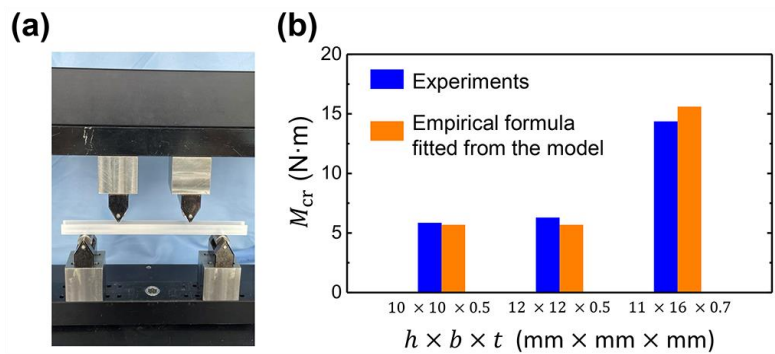

**Fig. 4 Experimental validation using four-point bending test. (a)** Image of the thin-walled U-shaped beam sample in 6063 aluminum alloy and the settings of experimental facilities. **(b)** Comparisons between the critical moments measured by experiments and estimated by the empirical formulas fitted from our model.

In addition to the pure bending focused in the modeling, diverse loading conditions exist in natural and engineering environments. This empirical formula can be modified and extended to estimate the critical loads for various loading conditions. Consider two thin-walled U-shaped cantilever beams, one with a concentrated force $F$ applied at the free end, and the other with a uniform transverse pressure $q$ applied at all cross sections (See Supplemental Material [28] for illustrations). These external loads cause the positive bending of U-shaped beams. Predictably, the buckling of sidewalls will concentrate on the region near the fixed ends of cantilever beams where the maximum bending moment is distributed, and also cause the asymmetric bending behaviors as the situation under pure bending. Similarly, we assume that, the buckling under these two loading conditions occurs when the maximum bending moment in the cross section reaches a critical value, and this critical moment is also proportional to $Et^3$. Then the critical force $F_{cr}$ and critical pressure $q_{cr}$ are ruled by $F_{cr} \propto Et^3/a$ and $q_{cr} \propto Et^3/a^2b$, respectively. Using the FEA results listed in Table S2 and S3 (See Supplemental Material [28] for detailed data), the coefficients of these two linear relationships are fitted as 0.84 and 1.94, respectively, with both $R^2$ of 0.99. It is verified that $F_{cr}$ and $q_{cr}$ show excellent linearity with respect to $Et^3/a$ and $Et^3/a^2b$, respectively. Therefore, we believe that our model can serve as a helpful reference for the analysis of similar problems under different loading conditions.

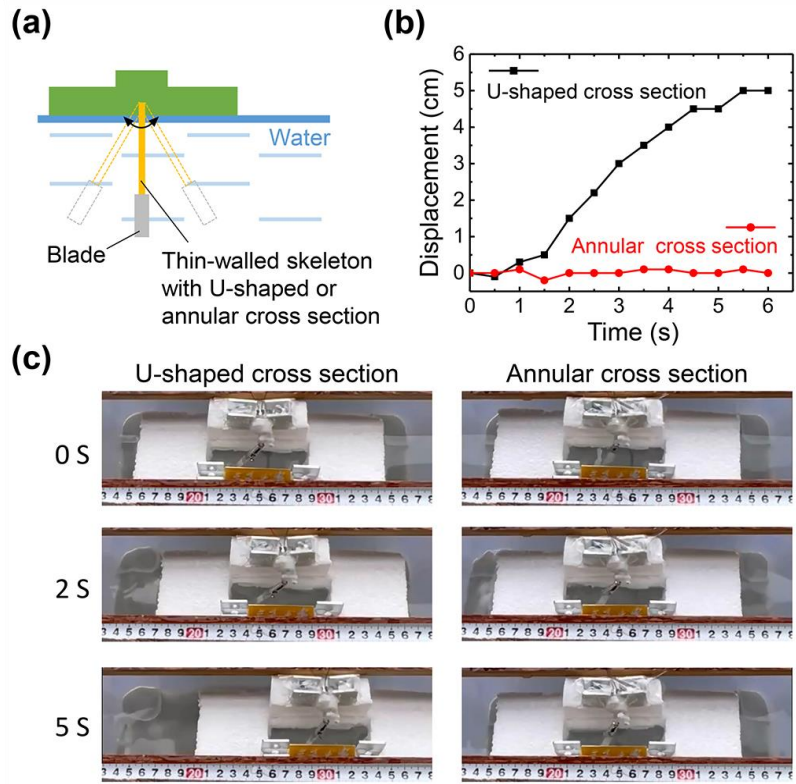

**Fig. 5 Enhanced locomotion performance brought by the design of thin-walled U-shaped beam structure. (a)** Prototype of a floating system propelled by a paddle that is assembled by a thin-walled skeleton and a blade. **(b)** Comparison between the displacements of floating systems equipped with skeletons in U-shaped and annular cross sections under the same actuation period. **(c)** Snapshots of the positions at different times.

Some explorations could be continued based on our findings. Further experimental studies are highly expected to design the "feet" or "wings" of soft robots into the thin-walled U-shaped beam structure for the verification of enhanced propulsion efficiency. Here, a primitive prototype is proposed as the schematic illustrated in Fig. 5a. We design a floating system propelled by a paddle that is assembled by a thin-walled skeleton with U-shaped cross section and a blade. The skeleton can be made by various responsive materials that are sensitive to light, temperature, magnet, etc. With the external input of cyclically switching stimuli, the skeleton will be bent back and forth to swing the paddle. The reaction force created by the water will make the floating system move. Comparing with cross sections in symmetric shapes, the U-shaped design will bring the asymmetric bending behaviors to the skeleton then produce more net propulsion. We quantitatively test the locomotion performance of two floating systems equipped with skeletons in U-shaped and annular cross sections. For a simple demonstration of the structure-induced effect, we replace the responsive materials with common polymers to manufacture the skeleton, and use an electric motor to actuate the swing of paddle. As plotted in Fig. 5b, the displacements under the same actuation period are compared. For the U-shaped cross section, the direction movement of floating system is effectively achieved. However, for the annular cross section, the floating system only slightly sways around the initial position. The difference can be clearly seen from the snapshots shown in Fig. 5c (See Supplemental Material [28] for the videos). We hope that this structure design would be integrated with the responsive materials to build more fantastic soft actuators/robots. In addition, more structures or devices with asymmetric mechanical responses could be inspired based on this beam structure for wider applications.

In summary, we have demonstrated the mechanism of the asymmetric bending behavior of thin-walled U-shaped beams, inspired by the morphology of slender leaves of many plants. It is found that this novel mechanical property is attributed to the buckling of sidewalls caused by the bending-induced compressive effect. A semi-analytical solution of the critical moment for the sidewall buckling is derived by a simple but efficient model based on the Euler-Bernoulli beam theory and Kirchhoff-Love thin plate theory. Using this model, the effects of structural geometries of beam on the buckling are evaluated. The critical moment is almost proportional to the cube of wall thickness, while nearly insensitive to the beam width, height and length. According to these correlations, a simplified expression of the critical moment is then concluded as a convenient empirical formula for engineering guidelines, which is further modified to fit different loading conditions. The theoretical foundations of our results are well validated by FEA simulations and experiments. Although this model follows some approximations and has some limitations, it still sheds light on the key mechanism of this asymmetric bending behavior and inspires further explorations on the design of high-performance soft actuators/robots or other structures with asymmetric mechanical responses.

We thank Andre Eccel Vellwock and Chengyu Wu for helpful discussions. This work was supported by the National Natural Science Foundation of China (Grant No. 11972226).

*Supplemental Material for*

# New perspectives on asymmetric bending behavior:
# A lesson learned from leaves

Anran Wei [1], Zhenbin Guo [2], Fenglin Guo [1] *

[1] School of Naval Architecture, Ocean and Civil Engineering (State Key Laboratory of Ocean Engineering), Shanghai Jiao Tong University, Shanghai 200240, China

[2] Institute of Semiconductor Manufacturing Research, Shenzhen University, Shenzhen 518060, China

* Corresponding author, E-mail: *flguo@sjtu.edu.cn*

Here we provide more details on theoretical derivations of effective elastic coefficients, results of finite element analysis simulations and illustrations of loading conditions for the model extension.

# I. Theoretical derivations of effective elastic coefficients $k_1$ and $k_2$

The U-shaped beam under positive bending is divided into two parts by the neutral surface. The part above the neutral surface is regarded as thin plates under compression with lower sides connected to the remained bottom part. The bottom part is also a thin-walled beam with a U-shaped cross section in the size of $b \times h_2$, as depicted in Fig. S1a. Take the unit beam length for analysis. Since the wall thickness $t$ is much smaller than $b$ and $h_2$, the bottom part can be approximated as a two-dimensional frame system $\overline{ABC}$, which is illustrated in Fig. S1b. The bars $\overline{AB}$ and $\overline{BC}$ are firmly joint at point $B$. At the point $A$, the displacement along $z$-direction and the rotation are prohibited to represent the symmetry of the bottom structure. At the point $C$, the displacement along $y$-direction is fixed to hold the static determinacy of the frame system, while no constraints are applied on the movement along $z$-direction and the rotation, allowing the deformation of the sidewalls. If a transverse force $F$ is applied at the point $C$, an out-of-plane displacement $\delta$ will be generated in the sidewall. The relationship between the $F$ and $\delta$ can be described by the elastic coefficient $k_1$ as $F = k_1 \delta$. Many methods can be found in the textbooks of structural mechanics for the solution of $k_1$. Based on the unit load method from the principle of virtual work, we have

$$\delta = \frac{1}{EI'} \left( \int_0^{h_2} F y^2 dy + \int_0^{b/2} F h_2^2 dz \right) \tag{A1}$$

Substituting $I' = t^3/12$ into Eq. (A1), the $k_1$ is calculated as

$$k_1 = \frac{E t^3}{4 h_2^3 + 6 h_2^2 b} \tag{A2}$$

Similarly, another elastic coefficient $k_2$ can be used to define the dependence of the rotation angle $\theta$ on the applied moment $M$ as $M = k_2\theta$ for the point $C$. Using the same method, the following relationship can be obtained as

$$\theta = \frac{1}{EI'}\left(\int_0^{h_2} M dy + \int_0^{b/2} M dz\right) \tag{A3}$$

Then the $k_2$ is given by

$$k_2 = \frac{Et^3}{12h_2 + 6b} \tag{A4}$$

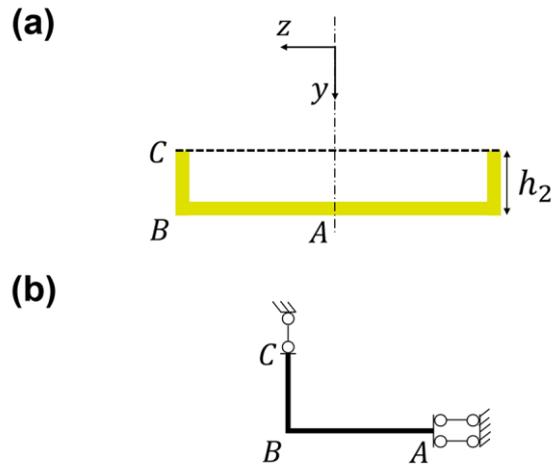

Fig. S1 Schematics of the derivations of elastic coefficients $k_1$ and $k_2$. For a U-shaped beam, the region below the neutral surface shown in **(a)** is modeled as a frame system $\overline{ABC}$ shown in **(b)**.

## II. Finite element analysis results for fitting the empirical formula of $M_{cr}$

| Cases | $v$ | $E$ (GPa) | $t$ (mm) | $a$ (mm) | $h$ (mm) | $b$ (mm) | $Et^3$ (N·m) | $M_{cr}$ (N·m) |
|---|---|---|---|---|---|---|---|---|
| 1 | 0.3 | 70 | 1 | 600 | 40 | 50 | 70.00 | 45.54 |
| 2 | 0.3 | 7 | 1 | 600 | 40 | 50 | 7.00 | 4.55 |
| 3 | 0.3 | 70 | 1.5 | 600 | 40 | 50 | 236.25 | 153.00 |
| 4 | 0.3 | 70 | 1 | 900 | 40 | 50 | 70.00 | 45.12 |
| 5 | 0.3 | 70 | 1 | 600 | 30 | 50 | 70.00 | 46.32 |
| 6 | 0.3 | 70 | 1 | 600 | 40 | 75 | 70.00 | 47.12 |

## III. Finite element analysis results for fitting the empirical formula of $F_{cr}$

| Cases | $v$ | $E$ (GPa) | $t$ (mm) | $a$ (mm) | $h$ (mm) | $b$ (mm) | $Et^3/a$ (N) | $F_{cr}$ (N) |
|---|---|---|---|---|---|---|---|---|
| 1 | 0.3 | 70 | 1 | 600 | 40 | 50 | 116.67 | 98.88 |
| 2 | 0.3 | 7 | 1 | 600 | 40 | 50 | 11.67 | 9.89 |
| 3 | 0.3 | 70 | 1.5 | 600 | 40 | 50 | 393.75 | 332.24 |
| 4 | 0.3 | 70 | 1 | 900 | 40 | 50 | 77.77 | 61.66 |
| 5 | 0.3 | 70 | 1 | 600 | 30 | 50 | 116.67 | 96.53 |
| 6 | 0.3 | 70 | 1 | 600 | 40 | 75 | 116.67 | 102.36 |

**IV. Finite element analysis results for fitting the empirical formula of $q_{cr}$**

| Cases | $v$ | $E$ (GPa) | $t$ (mm) | $a$ (mm) | $h$ (mm) | $b$ (mm) | $Et^3/a^2b$ (kPa) | $q_{cr}$ (kPa) |
|---|---|---|---|---|---|---|---|---|
| 1 | 0.3 | 70 | 1 | 600 | 40 | 50 | 3.89 | 7.60 |
| 2 | 0.3 | 7 | 1 | 600 | 40 | 50 | 0.39 | 0.76 |
| 3 | 0.3 | 70 | 1.5 | 600 | 40 | 50 | 13.13 | 25.55 |
| 4 | 0.3 | 70 | 1 | 900 | 40 | 50 | 1.73 | 3.07 |
| 5 | 0.3 | 70 | 1 | 600 | 30 | 50 | 3.89 | 7.27 |
| 6 | 0.3 | 70 | 1 | 600 | 40 | 75 | 2.59 | 5.25 |

## V. Illustrations of loading conditions for the model extension

Our model based on pure bending can be extended to estimate the critical loads of thin-walled U-shaped beams under other loading conditions. Here two cases that cause the positive bending are considered for the model extension, as illustrated in Fig. S2.

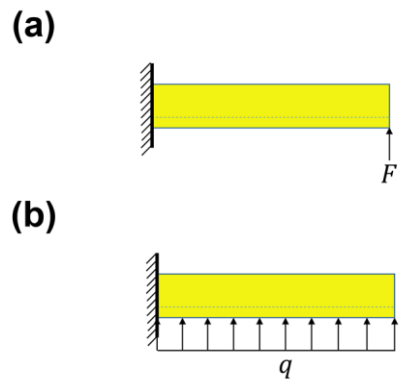

**Fig. S2 Some loading conditions that cause the positive bending of thin-walled U-shaped beams.** In **(a)** and **(b)**, the beams are placed with their U-shaped cross sections upward.